%% file: sabinacorr.tex
\documentstyle[12pt,chicagob]{article}
\input{defn}


\renewcommand{\H}{\Hcal}
\newenvironment{oldthm}[1]{\par\noindent{\bf Theorem #1:} \em \noindent}{\par}
\newenvironment{oldlem}[1]{\par\noindent{\bf Lemma #1:} \em \noindent}{\par}
\newenvironment{oldcor}[1]{\par\noindent{\bf Corollary #1:} \em \noindent}{\par}
\newenvironment{oldpro}[1]{\par\noindent{\bf Proposition #1:} \em \noindent}{\par}
\newcommand{\othm}[1]{\begin{oldthm}{\ref{#1}}}
\newcommand{\eothm}{\end{oldthm} \medskip}
\newcommand{\olem}[1]{\begin{oldlem}{\ref{#1}}}
\newcommand{\eolem}{\end{oldlem} \medskip}
\newcommand{\ocor}[1]{\begin{oldcor}{\ref{#1}}}
\newcommand{\eocor}{\end{oldcor} \medskip}
\newcommand{\opro}[1]{\begin{oldpro}{\ref{#1}}}
\newcommand{\eopro}{\end{oldpro} \medskip}
\begin{document}

\title{Expressing Security Properties Using Selective Interleaving Functions}
\author{Joseph Y. Halpern and Sabina Petride}
\maketitle
\begin{abstract}
McLean's notion of {\em Selective Interleaving Functions} ({\em SIF}s) is
perhaps the best-known attempt to construct a framework for expressing
various security properties.
We examine the expressive power of SIFs carefully.
We show that SIFs cannot capture {\em  nondeducibility on strategies} ({\em
NOS}).
We also prove that the set of security properties  expressed with SIFs
is not closed under conjunction, from which it
follows that {\em separability\/} is strictly stronger than double
generalized noninterference.
However, we show that if we generalize the notion of SIF in a natural way,
then NOS is expressible, and the set of security properties expressible by generalized SIFs is closed
  under conjunction.
\end{abstract}

\section{Introduction}
Trying to formalize what it means for a system to be secure
is a far from trivial task.  Many definitions of security have been
proposed, using quite different formalisms.  One intuition that many of
these definitions have tried to capture is that a system is secure if
no information flows from a higher-level user to a lower-level user
\cite{GM82}.
(From here on in, we just call these users {\em high} and {\em low},
respectively.)
This intuition, in turn, is captured by saying that,  given
their local observations, low users cannot rule out any possible
behavior of high users.  But even this intuition can be formalized
in a number of ways, depending on what we understand by ``high behavior''
and
on what kind of information we specifically want to protect.

Many current approaches to defining security
(for example, \cite{mclean90,McLean94,Wittbold&Johnson,mccullough87})
assume that high and low users
send input values to the system, which responds with output
values.  The ``system'' is then modeled as a set of
sequences ({\em traces}) of low/high  input and output values.
Various definitions of security then impose conditions on the set of
possible traces.

The following are some of the best-known definitions from the literature:
\begin{itemize}%
\item {\em Separability} (abbreviated SEP) \cite{McLean94} is one of the
most restrictive
definitions.  It requires that the system can be viewed as being
composed of two independent subsystems, corresponding to the low and high
users: every possible trace generated by the low user is compatible
with every trace produced by the high user.
While a separable system is certainly secure under any reasonable
definition of security, it is unrealistic to expect systems to be
separable in practice.  
Moreover,
not all interactions between high and low users may be seen
as a breach in the system's security.  After all, the main motivation behind
theories of information flow is to understand which types of such
interactions are admissible.
\item We can slightly relax separability by requiring only that the low
activity
be independent of the sequence of high inputs. The new property is called
{\em generalized noninterference\/} (GNI) \cite{mccullough87}.

\item
Traces are not  generated at random.  They usually come as a result of {\em
strategies}:
rules that stipulate the next input based on the history of input-output
values.
It has been argued that security really involves the low user not
finding out anything about the high user's strategy.  This notion
is captured by {\em nondeducibility on strategies\/} (NOS)
\cite{Wittbold&Johnson}.
\end{itemize}

Given all these different notions of security, it is helpful to have
a single unified framework in which to express them and compare their
relative strengths.
One attempt to do so was suggested by McLean
\citeyear{mclean90,McLean94}. 
McLean observed that most of the above security properties may
be expressed as {\em closure conditions on systems} (e.g. on sets of
traces): a system satisfies a given 
security property if for every pair of traces in the system there is a
trace in the system satisfying  
certain properties. This intuition is formalized by associating to a
security property a set $F$ of functions from pairs of traces to traces;
such a mapping from pairs of traces to traces is called a {\em selective
interleaving function} (SIF). 
A system $\Sigma$ is said to satisfy a security property if it is {\em
closed} under the associated set $F$ of  
SIFs, i.e., 
for all $\sigma_1, \sigma_2 \in \Sigma$ there is some $f \in F$ such that
$f(\sigma_1,\sigma_2) \in \Sigma$.
McLean focuses on some particularly natural sets of SIFs that he calls
{\em types}. To understand the notion of a type, we need to look more carefully at
the structure of traces.
Traces are assumed to be sequences of tuples of the form
(high input, low input, high output, low output).
A type consists of all SIFs that, given two traces as arguments, combine some
components from the first trace with some components from the second and that 
satisfy certain restrictions (for example, combining the high
input from the first trace and the low output from the second trace).

McLean shows that a number of security properties, including SEP and
GNI, can be represented by types in the sense that there exists a type $T$
such that a system $\Sigma$ has security property $S$ if and only if
$\Sigma$ is closed under type $T$.
He thus suggests that 
types
provide a reasonable framework in which to examine security properties.
Zakinthinos and Lee \citeyear{ZL97} point out
that, 
in their system model (which is slightly different from that used by
McLean---see Section~\ref{sec:ZL}), there are security
properties that cannot be expressed in terms of closure under types.
In this paper, we examine this question more carefully.

We show that NOS can {\em not} be represented
by types.
We also show that another natural property that we call {\em double
generalized noninterference (DGNI)\/} cannot be expressed either.
DGNI requires both that low activity is independent of the high
inputs and that high activity is independent of the low inputs.
The counterexample for DGNI actually proves the more general result
that security properties expressible by types are not closed under
conjunction.
More precisely, there are types $T_1$ and $T_2$ such that
for  no type $T$ is it the case that  a system is
closed under both $T_1$ and $T_2$ if and only if it is
closed under $T$.

These negative results are proved under the assumption that the only
sets of SIFs  are types.  If we allow more general sets of SIFs,
these results no longer hold.  NOS and DGNI
are all expressible; moreover, in the more general setting, we have
closure under conjunction.
However, considering closure under arbitrary sets of SIFs is arguably
not the most natural setting in which to examine security properties.
Moreover, it is far from clear that even this setting is as expressive
as we would like.

The rest of the paper is organized as follows.
Section~\ref{sec:review} reviews the formal definitions of the security
properties discussed above and McLean's SIF framework.
Section~\ref{sec:negative} contains
the negative results of the paper.  It shows that NOS and DGNI
can not be represented by types.
Section~\ref{sec:positive} shows that these negative results do not hold
if we consider closure under sets of SIFs more general than types.  In
fact, under the assumption that the set of traces is countable, this
framework
captures {\em all} security properties.
Section~\ref{sec:ZL} relates our results to those of
Zakinthinos and Lee \citeyear{ZL97}.  We conclude in
Section~\ref{sec:conc} with some discussion of the general issue of
representing security properties. 

\section{Security Properties and SIFs: A Review}\label{sec:review}
{\bf Notation}: Following McLean \citeyear{McLean94}, a trace $\sigma$
is a sequence of tuples of the form  
(high input, low input, high output, low output). 
We assume that we are given a set $\Sigma^*$ of traces 
(which McLean \citeyear{McLean94} calls the {\em trace space}).
Intuitively, $\Sigma^*$ is the set of all possible traces.

\dfn\label{system}
A {\em system\/} $\Sigma$ (in $\Sigma^*$) is a subset of $\Sigma^*$.
\edfn
Intuitively, $\Sigma$ is a collection of traces generated according to
some protocol or protocols. 
McLean implicitly assumes that traces are infinite. We allow traces to be finite or infinite (although we could equally 
well restrict to sets $\Sigma^*$ that have just finite or just infinite traces). Note that, because of the form of traces, the system is synchronous.

Let $2^{\Sigma^*}$ be the power set of $\Sigma^*$.
\dfn A {\em security property\/} $S$ (on $\Sigma^*$) is a predicate on
  $2^{\Sigma^*}$; that is, a security property is a set of systems in
$\Sigma^*$.
\edfn

 Intuitively,  $S$ picks out some systems in $\Sigma^*$
as the ``good'' systems, the ones that satisfy the property.
We may not want to allow an arbitrary set of systems to be a security
property.
However, we have not come up yet with any reasonable restrictions
on the sets of systems that count as security properties.
Interestingly, Zakinthinos and Lee \citeyear{ZL97} do put a
restriction on what counts as a security property.
We discuss their restriction in
Section~\ref{sec:ZL} and argue that it is not particularly well
motivated.
Note that our negative results consider
specific sets of systems that correspond to security properties that have
already been considered in the literature, so they should satisfy any
reasonable restrictions we may want to place on the definition.

\dfn
Given a trace $\sigma$, we denote by $\sigma|_L$ the {\em low view} of 
$\sigma$, the sequence consisting of (low input, low output) projection. We
similarly denote by $\sigma|_H$ the {\em high view} of $\sigma$, and by
${\sigma}|_{HI}$ the sequence consisting  just of the high inputs. 
\edfn
We can now formalize the notions of security
discussed in the Introduction.

\paragraph{Separability}\label{Sep}
 As mentioned before, SEP is a strong security requirement that
the low and high  events be
independent, meaning that any low view of a trace should be compatible with
any
high view of a trace.
Formally, a system $\Sigma$ satisfies SEP if
  $$\forall {\sigma}_1,{\sigma}_2 \in \Sigma,  \quad \exists
\sigma \in \Sigma \quad ( {\sigma}|_L={\sigma}_1|_L \: \wedge \:
{\sigma}|_H={\sigma}_2|_H ).$$
Thus, if $\Sigma$ satisfies SEP, then we can combine the low view of one
trace in $\Sigma$ and the high view of another trace in $\Sigma$ to
obtain a trace in $\Sigma$.
Notice that SEP is a closure condition on the set of traces, since for every pair of traces in $\Sigma$, there is a trace in $\Sigma$ with a specific property
(namely, the same low view as the first trace, and the same high view
as the second trace). 
\paragraph{GNI and DGNI}\label{GNI}
GNI is a weakening of SEP.  A system $\Sigma$ satisfies GNI if
the low view of one trace is compatible with the high input
view of any other trace; that is,
$$\forall {\sigma}_1, {\sigma}_2 \in \Sigma \  \ \exists \sigma \in
\Sigma \, ( {\sigma}|_L={\sigma}_1|_L \: \wedge \:
{\sigma}|_{HI}={\sigma}_2|_{HI}) .$$
As SEP, GNI is a closure condition on the set of traces.
Notice that, unlike SEP,  GNI places no
constraints  on the high output sequence  in $\sigma$.

A system $\Sigma$ satisfies {\em reverse GNI\/}
(RGNI)
 if, 
$$\forall {\sigma}_1, {\sigma}_2 \in \Sigma \ \  \exists
\sigma  \in \Sigma \ 
  ({\sigma}|_H={\sigma}_1|_H \: \wedge \:
{\sigma}|_{LI}={\sigma}_2|_{LI})  .$$
Again, RGNI is a closure condition on the set of traces.

A system $\Sigma$ satisfies {\em double GNI 
}(DGNI)
 if it satisfies
both GNI and reverse GNI.
Unlike the above properties, DGNI is not a closure condition on the set of traces; it is the conjunction of two such closure conditions.

Clearly SEP implies GNI and DGNI: given $\sigma_1$ and
$\sigma_2$, the
trace $\sigma$ guaranteed to exist by SEP satisfies all the properties
required for GNI and DGNI.  However, as we shall see, the converse
does not hold in general.

\paragraph{Nondeducibility on strategies}\label{NOS}
  Wittbold and Johnson~\citeyear{Wittbold&Johnson} pointed out that in
security it is often necessary to take into account the strategies being
used by low and high to generate the traces.
A {\em protocol for user $u$\/} determines the input that $u$
provides to the system as a function of $u$'s previous  input and
output values.
A protocol for the system determines the high and low output values as a
function of previous high and low inputs and outputs and the current
high and low inputs.

Protocols can be nondeterministic or probabilistic.
In this paper we  do not consider probabilistic protocols, since the
security conditions we consider are possibilistic (that is, they make no
mention of probabilities).
For the purposes of this discussion, assume that
the low user is following a fixed protocol $P_L$ and the system
is following a fixed protocol $P_S$. 
Let $\Hcal^*$ be the set of all possible high protocols.
If $H \in \Hcal^*$, 
let $\Sigma_H$ be the set of traces generated by running $(P_S,P_L,H)$.
If $\Hcal \subseteq \Hcal^*$, then define $\Sigma_{\Hcal} = \union_{H
\in \Hcal} \Sigma_H$.  (Note that this is not necessarily a
disjoint union.) 
Let $\cS_{\Hcal^*}$ consist of all systems of the form
$\Sigma_{\Hcal}$ for some $\Hcal \subseteq \Hcal^*$. 

\commentout{ 
We assume that $\Hcal^*$ is such that for  any two distinct subsets $\Hcal$
and $\Hcal'$ of $\Hcal^*$, we have $\Sigma_{\Hcal} \ne
\Sigma_{\Hcal'}$.  It is easy to see that this is equivalent to the
assumption that for any protocol $H \in \Hcal^*$ and subset $\Hcal
\subseteq \Hcal^*$ such that $H \notin \Hcal$, we have that $\Sigma_H -
\Sigma_{\Hcal} \ne \emptyset$.  That is, there is a trace generated by
$H$ that is not generated by any protocol in $\Hcal$.
This assumption ensures that there is an injective mapping from $\Hcal$
to $\Sigma_{\Hcal}$, and thus makes it easier to state our results.

With this background, we can define NOS.
Note that it is defined only in systems of the form $\Sigma_{\cal H}$.
The system $\Sigma_{\cal H}$ satisfies NOS if
$$\forall \sigma \in \Sigma_{\cal H} \; \forall H \in {\cal H} \;
\exists {\sigma}^H
\in {\Sigma}_H \; ( {\sigma}^H|_L={\sigma}|_L ) .$$
Thus,
for every trace $\sigma \in \Sigma_{\Hcal}$ and every high
strategy $H \in \Hcal $, there must be a trace ${\sigma}^H \in \Sigma_{\Hcal}$
where the high user runs $H$ and the low user's view is the same as in $\sigma$.
}
With this background, we can define NOS.
The system $\Sigma_{\cal H}$ satisfies NOS if
$$\forall \sigma \in \Sigma_{\cal H} \; \forall H \in {\cal H} \;
\exists {\sigma}^H
\in {\Sigma}_H \; ( {\sigma}^H|_L={\sigma}|_L ) .$$
Thus, for every trace $\sigma \in \Sigma_{\Hcal}$ and every high
strategy $H \in \Hcal $, there must be a trace ${\sigma}^H \in \Sigma_{\Hcal}$
where the high user runs $H$ and the low user's view is the same as in $\sigma$.
Note that NOS is defined only for systems of the form $\Sigma_{\cal H}$.

For the definition above to make sense, it must be the case that 
two sets $\Hcal$ and $\Hcal'$ of protocols generate the same set of traces,
i.e., if  $\Sigma_{\Hcal}=\Sigma_{\Hcal'}$, then
$\Sigma_{\Hcal}$ satisfies NOS if and only if $\Sigma_{\Hcal'}$ satisfies NOS.
One way to ensure this is by focusing on sets of strategies $\Hcal^*$
such that there is an injective 
mapping from $\Hcal$ to $\Sigma_{\Hcal}$; in other words, 
if $\Hcal$ and $\Hcal'$ are distinct subsets of $\Hcal^*$, then we have
$\Sigma_{\Hcal}\neq \Sigma_{\Hcal'}$. 
This is equivalent to requiring that for any protocol $H \in
\Hcal^*$ and subset 
$\Hcal \subseteq \Hcal^*$ such that $H \not \in \Hcal$, we have $\Sigma_H - \Sigma_{\Hcal}\neq \emptyset$.
To see why this the case, 
suppose first that if $\Hcal \ne \Hcal'$, then $\Sigma_{\Hcal}\neq
\Sigma_{\Hcal'}$.
Let $\Hcal \subseteq \Hcal^*$ and $H\in \Hcal^*$ such that 
$H \not \in \Hcal$. Then we can simply take $\Hcal'=\lbrace H\rbrace
\cup \Hcal$ and since 
$\Hcal'\neq \Hcal$, we can apply the hypothesis and deduce that $\Sigma_{\Hcal'}\neq \Sigma_{\Hcal}$, or
equivalently, $\Sigma_H \cup \Sigma_{\Hcal}\neq \Sigma_{\Hcal}$. 
This means that $\Sigma_H - \Sigma_{\Hcal}\neq \emptyset$. 
For the converse, suppose that $\Sigma_H - \Sigma_{\Hcal}\neq \emptyset$
for all $H$ and $\Hcal$ such that $H \not \in \Hcal$. 
Let $\Hcal$ and $\Hcal'$ be two distinct subsets of $\Hcal^*$; since
$\Hcal \neq \Hcal'$,  
either $\Hcal - \Hcal'\neq \emptyset$, or $\Hcal' - \Hcal\neq
\emptyset$. Without loss of generality, we can assume 
that we are in the first case, and let $H$ be a strategy in 
$\Hcal - \Hcal'$.  By assumption,
$\Sigma_H - \Sigma_{\Hcal'}\neq \emptyset$.
Since $H \in \Hcal$, it follows that
$\Sigma_H \subseteq \Sigma_{\Hcal}$, and so $\Sigma_{\Hcal}- \Sigma_{\Hcal'}\neq \emptyset$;
in particular,  $\Sigma_{\Hcal}\neq \Sigma_{\Hcal'}$.
In short, for the definition of NOS to make sense, it suffices to 
 assume that for any strategy $H$ and set $\Hcal$ such that $H\not \in \Hcal$,
there is a trace generated by $H$ that is not generated by any protocol in $\Hcal$.
For the rest of the paper, we make this assumption when dealing with NOS.
It is interesting to notice that NOS is not a closure condition on the
set of traces, which suggests a 
different nature of NOS from SEP or GNI; this intuition will be
formalized in Theorem~\ref{thm:notNOS}. 

These security properties are related.
\pro\label{pro:relationships} Let $\Sigma$ be a system and let $\Hcal
\subseteq \Hcal^*$.
\begin{itemize}
\item[(a)] If $\Sigma$ satisfies SEP, then it satisfies DGNI.
\item[(b)] If $\Sigma$ satisfies DGNI, then it satisfies GNI.
\item[(c)] If $\Sigma_{\Hcal}$ satisfies SEP, then it satisfies
NOS.
\end{itemize}
\epro

\prf
Parts (a) and (b) are almost immediate from the definitions. For part
(c), suppose that system  
$\Sigma_{\Hcal}$  satisfies SEP, $\sigma \in \Sigma_{\Hcal}$, 
and $H \in {\cal H}$.  Choose $\sigma^H \in \Sigma_H$.
(There must always be at least one trace generated by running
$(P_S,P_L,H)$, so $\Sigma_H \ne 
\emptyset$.)
By SEP, there exists some $\sigma' \in \Sigma_{\Hcal}$ such that
${\sigma}'|_L={\sigma}|_L$ and ${\sigma}'|_H={\sigma}^H|_H$.
Since the inputs determined by $H$ at time $k+1$ depend only on the
sequence of $H$'s input and output values up to and including time $k$,
it immediately follows that $\sigma' \in \Sigma_H$. 
 Thus, $\Sigma_{\Hcal}$ satisfies NOS.
\eprf

The converses to (a), (b), and (c) do not hold in general,
as the following examples show.

\xam\label{xam:DGNInotSEP}
Let $\Sigma_{DGNI}$ consist of the 15 traces of the form
(As usual, we use the notation $(x_1,x_2,x_3,x_4)^\omega$ to denote the
trace where $(x_1,x_2,x_3,x_4)$ repeats forever.)
It is easy to see that this system does not satisfy
SEP (for example, $(0,0,0,0)^\omega$ and $(1,1,1,1)^\omega$ are in
$\Sigma_{DGNI}$, but $(1,0,1,0)^\omega$ is not), but does satisfy DGNI.
\exam

\xam\label{xam:GNInotDGNI}
Consider the system $\Sigma_{GNI} = \{\sigma_1, \sigma_2, \sigma_3,
\sigma_4\}$, where
$\sigma_1 = (1,0,1,0)^\omega$, 
$\sigma_2 = (1,1,0,1)^\omega$, 
$\sigma_3 = (0,0,0,0)^\omega$, and
$\sigma_4 = (0,1,1,1)^\omega$.  
It is easy to check that $\Sigma_{GNI}$
satisfies GNI, but it does not satisfy DGNI, since there is no trace
$\sigma \in \Sigma_{GNI}$ such that $\sigma|_H = \sigma_4|_H$ and
$\sigma|_{LI} = \sigma_3|_{LI}$.
\exam

\xam\label{xam:NOSnotSEP}
Let ${\cal H}^*$ consist of one protocol $H$; according to $H$, the high
user
first inputs 0 and then, at each step, inputs the previous low input value.
Let $P_L$'s protocol be such that, initially, the low user
nondeterministically chooses either 0 or 1, and then inputs that value
at every step. Finally, let the system protocol be such that the low
output and high output agree with the low input.  The system
$\Sigma_{NOS}$ generated by this protocol consists of two traces:
$(0,1,1,1){(1,1,1,1)}^\omega$ and $(0,0,0,0)^\omega$.  Since ${\cal H}^*$ consists of only
one protocol, $\Sigma_{NOS}$ trivially satisfies NOS.  It is also
immediate that $\Sigma_{NOS}$ does not satisfy SEP, since
$(0,1,0,1)^\omega$ is not in $\Sigma_{NOS}$.
\exam
One common trait of the majority of the above security properties is their correspondence to 
closure conditions on sets of traces (e.g. on systems): a system $\Sigma$ satisfies a 
security property if some closure condition on $\Sigma$ holds. One way to formalize this
approach is to associate to each security property a set $F$ of functions from pairs of traces 
to traces. 
\dfn\label{SIF}
A {\em SIF\/} (on $\Sigma^*$) is a partial function $f: \Sigma^* \times
\Sigma^* \rightarrow \Sigma^*$.  That is, a SIF takes two traces and (if
defined) returns a trace. 
\edfn
Our notion of SIF slightly extends McLean's by allowing partial
functions; this is 
convenient for the positive results in Section~\ref{sec:positive}.

\dfn\label{closureSIFs}
A system $\Sigma$ is {\em closed\/} under a set $F$ of SIFs if,
for all ${\sigma}_1,{\sigma}_2 \in \Sigma$, there exists some $f \in F$
such that ($f(\sigma_1,\sigma_2)$ is defined and)
$f({\sigma}_1,{\sigma}_2) \in \Sigma$.%
\footnote{We remark that McLean \citeyear{McLean94} actually does not make
it clear if the choice of $f$ can depend on the pair of traces, although
it seems that it can.  In  any case, in our positive results, we show that
can take the $f$ to depend only on the system, not the traces.  Indeed,
in the framework of Section~\ref{sec:positive}, the two choices lead to
equivalent definitions.}
\edfn

Of particular interest are certain sets of SIFs called {\em
types}.%
\footnote{
We remark that McLean \citeyear{McLean94} occasionally interchanges the terms
{\it function} and {\it type}.  For example, when he says that
a system is closed under a {\it
function}, what is meant is that the system is actually closed under the
{\it type} of the function (that is, under the set of functions of a
particular type).  We have tried to be careful in our usage here.}
\dfn
A SIF $f$ has {\em type\/} $\langle (in^H:in^L),(out^H:out^L)\rangle$,
where $in^H, in^L, out^H, out^L \in \{0,1,2\}$,
if $f$ is
total and 
$\sigma_3 = f(\sigma_1, \sigma_2)$ 
satisfies the following constraints:
\begin{itemize}
\item If $in^H = 1$, then $\sigma_3|_{HI} = \sigma_1|_{HI}$: the high
inputs of $f(\sigma_1,\sigma_2)$ is the same as the high input of
$\sigma_1$.
\item If $in^H = 2$, then $\sigma_3|_{HI} = \sigma_2|_{HI}$: the high
inputs of $f(\sigma_1,\sigma_2)$ is the same as the high input of
$\sigma_2$.
\item If $in^H = 0$, then there are no constraints on $\sigma_3|_{HI}$.
\end{itemize}
There are 9 other similar clauses, depending on the value of the
other components in the tuple.
\edfn
Thus, for example, if $f$ has type $\langle (1:2), (0:2)\rangle$ and $f({\sigma}_1,{\sigma}_2)=\tau$, then 
$\tau \in {\Sigma}^*$,
${\tau}|_{HI}={{\sigma}_1}|_{HI}$ (the high input views of $\tau$ and ${\sigma}_1$ are identical),
${\tau}|_{LI}={{\sigma}_2}|_{LI}$ (the low input views of $\tau$ and ${\sigma}_2$ are identical),
there is no restriction on the high output view ${\tau}|_{HO}$ of $\tau$, and ${\tau}|_{LO}={{\sigma}_2}|_{LO}$ (the low output views of $\tau$ and ${\sigma}_2$ are identical).

Let $T_{\<(i_1,i_2),(j_1:j_2)\>}$ consist of all SIFs of type
 $\<(i_1,i_2),(j_1,:j_2)\>$.  Note that if none of $i_1$, $i_2$ ,$j_1$,
or $j_2$ is 0, then $T_{\<(i_1,i_2),(j_1:j_2)\>}$ is a singleton set.

If there is a single high  user and a single low  user (as we
have been
assuming here) there are 81 possible types.  (Not all these types are
distinct, as we shall see.)
Since a type is just a set of SIFs, it makes sense to talk about a
system being closed under a type, using our earlier definition.

\dfn Let ${\cal S'}$ be a set of systems (i.e., subsets of $\Sigma^*$)
and let $\cS$ be a security property.  (Recall that a security property
is also a set of systems.)  A type $T$
{\em $\cS'$-represents\/} a security property $\cS$ with respect to
$\cS'$ if, for all systems $\Sigma \in \cS'$,  $\Sigma \in \cS$ if and
only if $\Sigma$ is closed under type $T$.
\edfn

The reason that we allow the generality of representation with respect
to a set $\cS'$ of systems is that, in the case of  NOS, we are
interested only in systems in $\cS_{\Hcal^*}$ (that is, systems of
the form $\Sigma_{\Hcal}$ for some $\Hcal \subseteq \Hcal^*$).  
Let $\cS^*$ denote the set of all subsets of $\Sigma^*$.
McLean   shows that SEP and GNI can both be
represented by types.
\pro \label{types for sep and GNI}
{\rm \cite{McLean94}}
\begin{itemize}
\item[(a)] SEP is $\cS^*$-represented by the type $T_{\langle
(1:2),(1:2)\rangle}$.
\item[(b)] GNI is $\cS^*$-represented by the
type $T_{\langle (1:2),(0:2)\rangle}$.
\end{itemize}
\epro

\noindent
McLean \citeyear{McLean94} also shows that other security properties,
such 
as {\em noninference\/} \cite{O'Halloran}, {\em generalized
noninference\/} \cite{McLean94}, and {\em noninterference\/}
\cite{GM82}, are represented by types. 
\commentout{
NF prevents 
any information flow from the high to the low level by requiring that a
low user cannot infer that high inputs or outputs have
occurred; this is captured formally by ensuring that the low view of
any trace is also a trace in the system. However, {\em NF} prevents
certain kinds of flow from low to high level, limitation which is
avoided by {\em GNF} by demanding that only high inputs cannot be
inferred by a low level user. {\em NI} is the requirement that
high-level users cannot affect what low-level users can see. For
deterministic systems, {\em NI} is equivalent to {\em NF}, assuming that
high level output cannot be generated when there is no high level
input. For nondeterministic systems, {\em NF} is more general than {\em
NI}.   
\begin{figure}
\begin{picture}(200,200)(10,10)
\put(100,190){SEP $T_{\langle (1:2),(1:2)\rangle}$}
\put(110,189){\vector(1,-1){35}}
\put(110,189){\vector(-1,-3){42}}
\put(55,50){NF}
\put(150,150){NI}
\put(155,149){\vector(0,-1){40}}
\put(150,100){DGNI}
\put(155,99){\vector(0,-1){40}}
\put(150,50){GNI $T_{\langle (1:2),(0:2)\rangle}$}
\put(150,49){\vector(-2,-1){30}}
\put(90,30){GNF}
\put(58,49){\vector(2,-1){30}}
\end{picture}
\caption{Partial order of security properties}
\end{figure}
}
\section{Types are Insufficiently Expressive}\label{sec:negative}

Although McLean did show that a number of security properties of
interest can be represented by types, given that there are only 81 types,
it is perhaps not surprising that there should be some interesting
security properties that are not  representable by any type.  In this
section, we prove the two negative results discussed in the
introduction: that neither
NOS nor DGNI are representable by types, and that
the properties representable by types are not closed under conjunction.
We also show that the properties represented by types are not closed
under disjunction either.

\thm\label{thm:notNOS} NOS is not $\cS_{\Hcal^*}$-representable by a type.
\ethm

\thm\label{thm:DGNI} DGNI is not $\cS^*$-representable by a type. \ethm

Since there are only $3^4=81$ possible types,
we can prove both Theorem~\ref{thm:notNOS} and~\ref{thm:DGNI} by
checking each of these types.   We make a number of observations that
allow us to significantly reduce the number of types that need to be
checked, making it a manageable problem.
We leave details to the appendix.

Theorem~\ref{thm:DGNI} is actually an instance of a more general 
result.

\dfn
A set $\P$ of security properties is {\em closed under conjunction\/} if
$\cS_1, \cS_2 \in \P$ implies that $\cS_1 \inter \cS_2 \in \P$.
Similarly, $\P$ is {\em closed under disjunction\/} if for all $\cS_1,
\cS_2 \in \P$ implies that $\cS_1 \union \cS_2 \in \P$.
 \edfn
Closure under conjunction seems like a natural requirement for
security properties. We may be interested in systems that satisfy both security
property $\cS_1$ and security property $\cS_2$.  Closure under disjunction
may also be of interest; that is, we may  investigate a system that
satisfies either one of properties $\cS_1$ or $\cS_2$.

\cor The set of security properties representable by types is not closed
under conjunction.
\ecor
\prf GNI and reverse GNI
are representable by types, but the
security property resulting from their conjunction (DGNI) is not
representable by types. \eprf

\thm\label{disjTypes} The set of security properties representable by
types is not closed under disjunction.
\ethm

\prf See the appendix.
\eprf

\section{Representation by SIFs} \label{sec:positive}
The definition of closure under a set $F$ of SIFs makes sense for
arbitrary sets $F$, not just for types.  Thus, just as for types, we can say
that a set $F$ of SIFs {\em $\cS'$-represents\/} a security property
$\cS$ if, for all systems $\Sigma \in \cS'$, $\Sigma \in \cS$ if and only if
$\Sigma$ is closed
under $F$.  In this section we show that, if we consider arbitrary sets
of SIFs rather than types, the negative results of the previous section
no longer hold.
More specifically, we prove that NOS is representable by SIFs and that
the set of security 
properties representable by SIFs is closed under conjunction and disjunction. 
Furthermore, under certain assumptions (that are satisfied by most  systems 
of interest), 
we show that {\em every\/}
security property can be represented by SIFs. However, 
the representation is rather convoluted, and requires understanding what
set of systems satisfy the property.  This negates the whole point of
using the approach to describe properties.  If we already know what
systems satisfy the security property, we can just work with that set
directly.  
However, we show that a more uniform way of representing security
properties can be obtain by allowing
{\em generalized SIFs\/} that associate with each pair of traces a {\em
set} of traces. 
We start by showing that NOS is representable by SIFs.

\thm\label{NOS&SIFs}
NOS is $\cS_{\Hcal^*}$-representable by SIFs.
\ethm

\prf
We must find a set $F$ of SIFs such that  a system $\Sigma \in
\cS_{\Hcal^*}$
satisfies NOS if and only if it closed under $F$.
Given a protocol
$H \in \Hcal^*$ and a trace $\sigma \in {\Sigma}_H$,
let $f_{H,\sigma}(\sigma_1,\sigma_2)$ be the trace $\sigma$ if
${\sigma}|_{L} ={\sigma_1}|_{L}$ and $\sigma_2 \in {\Sigma}_H$,
and undefined, otherwise. (Recall that we allow partial
functions.)%
\footnote{
If we restrict to systems
$\Sigma^*$ and sets $\H^*$ such that there is some trace $\sigma_0 \notin
\union_{H \in \H^*} \Sigma_H$, then $\cS_{\H^*}$ is representable by
total SIFs.  The proof is essentially the same as that given for
Theorem~\ref{NOS&SIFs}, but rather than taking $f(\sigma_1,\sigma_2)$ to be
undefined in the proof, we take $f(\sigma_1,\sigma_2) = \sigma_0$.
It is then a matter of taste whether it is more reasonable to consider
partial SIFs or to assume that there are traces that cannot be generated
by any protocol.}
Let $F$ be the set of all such functions.
It is easy to show that  if $\Sigma_{\Hcal}$ satisfies NOS, then it is
closed
under $F$.  Now suppose that $\Sigma_{\Hcal}$ is closed under $F$.
Given
$\sigma \in \Sigma_{\Hcal}$ and $H \in \Hcal$, as we mentioned earlier,
by our assumption that $\Sigma_{\Hcal} \ne \Sigma_{\Hcal'}$ for any
$\Hcal'$ (in particular, for $\Hcal' = \Hcal- \{H\}$),
there must be a trace $\sigma^H$ generated by $H$ that is not generated
by any other protocol in $\Hcal$.  Since $\Sigma_{\Hcal}$ is closed under
$\cal
F$,
there is a function $f_{H',{\sigma'}}$ in $F$, such that
$f_{H',\sigma'}(\sigma,\sigma^H) \in \Sigma_{\Hcal}$. Then
$f_{H',\sigma'}(\sigma,\sigma^H)=\sigma'$, and ${\sigma'}|_L={\sigma}|_L$,
${\sigma'} \in {\Sigma}_H$. By definition of ${\sigma}^H$, it must be
the case that $H'=H$, so $\sigma' \in {\Sigma}_H$ and
${\sigma'}|_L={\sigma}|_L$. Thus, ${\Sigma}_{\Hcal}$ satisfies NOS.
\eprf

The following result is also easy to see.

\pro The security properties $\cS^*$-representable by SIFs
is closed under disjunction. \epro

\prf Suppose that $\cS_1$ is represented by $F_1$ and $\cS_2$ is
represented by $F_2$.  Then $\cS_1 \union \cS_2$ is represented by $F_1
\union F_2$. \eprf

These results show that allowing arbitrary SIFs gives much more
expressive power than just considering types.
Exactly how expressive are they?
As we now show, they are quite expressive:~if
$\Sigma^*$ is countable, then {\em every\/} security property is
representable by SIFs.
This already means that for many systems of interest, all security
properties are expressible with SIFs.
For example, if the underlying protocols being
represented by $\Sigma^*$ 
all terminate,
and there are only countably many of
them, then $\Sigma^*$ will be countable.  But if we allow nonterminating
protocols that, for example, nondeterministically output either 0 or 1
at every step, then the set of traces will be uncountable.
However, we can extend the result to uncountable sets, provided that
they are not ``unreasonable''.

Say that a set $\cS'$ of systems is {\em countably
generated\/} if for all $\Sigma \in \cS'$, there exists a countable set
$\Sigma_c$ of traces in $\Sigma$ such that if $\Sigma' \in \cS'$ and
$\Sigma' \subset \Sigma$, then there is a trace $\sigma \in \Sigma_c -
\Sigma'$.  Clearly if $\Sigma^*$ is countable, then any security
property on $\Sigma^*$ is countably generated.  (Just take $\Sigma_c =
\Sigma$.)  But the notion of countable generation also applies to
interesting possible uncountable systems.
Given a trace $\sigma$, let $\sigma_{1:n}$ be the prefix of $\sigma$ of
length $n$; if $\sigma$ is finite and has length less than $n$, then
$\sigma_{1:n} = \sigma$.
A set $\Sigma \subseteq \Sigma^*$ of traces is {\em limit closed\/}
\cite{Emerson} if
for every  $\sigma \in \Sigma^*$ and  for all $n \in N$ such that there
exists
$\sigma' \in \Sigma$ with $\sigma_{1:n} = \sigma'_{1:n}$, it is the case
that $\sigma \in
\Sigma$.
  Intuitively, $\Sigma$ is limit closed if, whenever it contains
every prefix of a trace $\sigma$, it also contains $\sigma$.

\lem If $\cS'$ consists only of limit-closed sets of traces, and the set
of possible inputs and outputs is countable, then $\cS'$
is countably generated. \elem

\prf Given $\Sigma$, let $A$ consist of all the prefixes of traces in
$\Sigma$.  Since the set of all inputs and outputs is countable, $A$
must be a countable set.  Let $\Sigma_f$ be a subset of $\Sigma$ such
that for each prefix $\tau$ of length $n$ in $A$, there exists a trace
$\sigma \in \Sigma_f$ such that $\sigma_n = \tau$.  Clearly we can take
$\Sigma_f$ to be countable.  Now let $\Sigma' \subset \Sigma$ be such that
$\Sigma' \in \cS'$, and let
$A'$ consist of all prefixes of traces in $\Sigma'$.  If $A = A'$, then
an easy argument shows that,
by limit closure, we must have $\Sigma = \Sigma'$.  Thus, there must be
some prefix $\tau$ in $A$ with no extension in $\Sigma'$.  By
construction of $\Sigma_f$, there is some trace $\sigma$ extending
$\tau$ in $\Sigma_f$.  Clearly, $\sigma \in \Sigma_f - \Sigma'$.
\eprf

Limit closure is a natural condition that arises frequently in
practice.  In particular, $\Sigma_H$ is limit closed.
Thus, $\cS_{\Hcal^*}$ is countably generated, even if the set of traces
in $\Sigma_{\Hcal^*}$ is uncountable.  In light of this, a good case can
be made that we are interested in $\cS'$-representability only for sets
$\cS'$ that are countably generated.

\thm\label{thm:countablegen} If $\cS'$ is countably generated, then all
security property are $\cS'$-representable by SIFs.
\ethm

\prf Suppose that $\Sigma \in \cS'$.  We show that there exists a SIF
$f_{\Sigma}$ such that $\Sigma$ is the only set in $\cS'$ that is closed
under $f_{\Sigma}$. It follows that the security property $\cS$ is
$\cS'$-representable
by the set of SIFs $\{f_{\Sigma}: \Sigma \in \cS\}$.

Since $\Sigma$ is in $\cS'$ and $\cS'$ is countably generated, there is a
countable subset ${\Sigma}_c$ of $\Sigma$
with the properties from the definition.
We take $f_{\Sigma}(\sigma,\sigma')$ to be undefined if at least one of
$\sigma$ and $\sigma'$ is not in $\Sigma$.
If both $\sigma$ and $\sigma'$ are in $\Sigma$, but only one of them is in
${\Sigma}_c$, then we take
$f_{\Sigma}(\sigma,\sigma')$ to be exactly the trace in ${\Sigma}_c$. If
none of the traces is in ${\Sigma}_c$, then
choose some trace ${\sigma}_c$ in ${\Sigma}_c$ and let it be equal to
$f_{\Sigma}(\sigma,\sigma')$.
There is one case left: both traces $\sigma$ and $\sigma'$ are in
${\Sigma}_c$.

Since ${\Sigma}_c$ is countable,
it is either finite or countably infinite.  If it is infinite, then
without loss of generality 
it has the form
$\{{\sigma}_k
| k \in \Z\}$. Then   $\sigma={\sigma}_i$ and  $\sigma'={\sigma}_j$ for some
$i$ and $j$. Let  $f_{\Sigma}({\sigma}_i,{\sigma}_j)={\sigma}_{i+1}$ if $j$
even, and
${\sigma}_{i-1}$ if $j$ odd. It is easy to see that $\Sigma$ is closed
under $f_{\Sigma}$. Suppose now that $\Sigma'$ in $\cS'$ is closed under
$f_{\Sigma}$ too. Then it must be the case that $\Sigma' \subseteq
\Sigma$. Suppose that $\Sigma \ne \Sigma'$.
By definition, there is some trace in ${\Sigma}_c$
that is not in $\Sigma'$.
Thus, there is some $i$ such that ${\sigma}_i \in
\Sigma'$, but at least one of ${\sigma}_{i-1}$ or ${\sigma}_{i+1}$ is not in
$\Sigma'$. Suppose that $\sigma_{i-1} \notin \Sigma'$.  If $i$ is odd, then
${\sigma}_{i-1}=f_{\Sigma}({\sigma}_i,{\sigma}_i)$, and since $\Sigma'$
closed under $f_{\Sigma}$ and ${\sigma}_i \in {\Sigma'}$, then
${\sigma}_{i-1}$ must be in $\Sigma'$, which contradicts our
supposition. If $i$ is even, then
$f_{\Sigma}({\sigma}_i,{\sigma}_i)={\sigma}_{i+1}$, so ${\sigma}_{i+1}$
must be in $\Sigma'$, and so
$f_{\Sigma}({\sigma}_i,{\sigma}_{i+1})={\sigma}_{i-1}$ is also in
$\Sigma'$, which is again a contradiction.
The argument if ${\sigma}_i \in
{\Sigma}'$, but ${\sigma}_{i+1} \not\in {\Sigma}'$ is similar, and left
to the reader.

If ${\Sigma}_c$ if finite, then we can write it as $\{
{\sigma}_1, \ldots ,{\sigma}_k \}$ for some $k$.
The proof is essentially the same, except that $i+1$ or $i-1$ are now
modulo  $k$.
\eprf

Although Theorem~\ref{thm:countablegen} shows that essentially every
security property can be represented by SIFs,
the representation is not terribly interesting.
The proof requires one to work backwards from an explicit
representation of the security property as a set of systems to the SIF.
To the extent that SIFs are going to be a useful tool for representing
security properties, then there should be a more uniform way of
representing security properties.  For example, the representation of
GNI or even NOS is essentially the same, independent of $\Sigma^*$.
We do not know if there is a uniform way of representing, say, DGNI
using SIFs, although it follows from Theorem~\ref{thm:countablegen} that
it can be represented in essentially all cases of interest.

Interestingly,
by somewhat extending the notion of SIF, we can give a more uniform
definition of DGNI, as well as proving closure under conjunction.  The
idea is to allow a SIF to associate to all pairs of traces not necessarily a
single trace, but a
{\em set} of traces.
\dfn\label{def: gen-sifs}
A {\em generalized SIF\/} is  a partial function from 
${\Sigma}^* \times {\Sigma}^*$ to $2^{{\Sigma}^*}$.
\edfn

Clearly if we restrict to functions whose values are
singletons, then we get SIFs as defined earlier.
Thus, Theorems~\ref{NOS&SIFs} and~\ref{thm:countablegen} continue to
hold in the extended framework.
But it is easy to see that the set of security properties representable by
generalized SIFs is closed under conjunction.
\pro The security properties $\cS^*$-representable by generalized SIFs
is closed under conjunction and disjunction. \epro
\prf
Suppose that ${\cal S}_1$ is ${\cal
S}^*$-representable by $F_1$, and ${\cal S}_2$ is ${\cal
S}^*$-representable by $F_2$, where $F_1$ and $F_2$
sets of generalized SIFs. For each $f \in {F}_1$ and $g \in {\cal
F}_2$, define $[f,g]({\sigma}_1,{\sigma}_2)$ to be undefined if either
$f({\sigma}_1,{\sigma}_2)$ or $g({\sigma}_1,{\sigma}_2)$ is undefined, and
$f({\sigma}_1,{\sigma}_2) \union g({\sigma}_1,{\sigma}_2)$
otherwise. Let $ F=\{ [f,g] \, : \, f \in {F}_1, g \in {\cal
F}_2\}$. It is easy to show that if $\Sigma \in {\cal S}_1 \inter {\cal
S}_2$, then $\Sigma$ is closed under $F$. Suppose now that $\Sigma$
is closed under $F$. Then
for all ${\sigma}_1,{\sigma}_2 \in
\Sigma$, there is some function $[f,g] \in {F}$ such that
$[f,g]({\sigma}_1,{\sigma}_2) \in \Sigma$. That means that
$f({\sigma}_1,{\sigma}_2)$ and $g({\sigma}_1,{\sigma}_2)$ are both
defined, and since their union is in $\Sigma$, each of then is a subset
of $\Sigma$. So $\Sigma$ is closed under $F_1$ and $F_2$;
that is, $\Sigma \in {\cal S}_1 \inter {\cal S}_2$.  Thus, we have
closure under conjunction.  The argument for closure under disjunction
is identical to that for SIFs.
\eprf

\cor DGNI is $\cS^*$-representable by generalized SIFs. \ecor

\section{Related Approaches}\label{sec:ZL}
Zakinthinos and Lee \citeyear{ZL97} (ZL from now on) also
consider the question of expressing security properties, although
their approach is slightly different from McLean's. 
They work in an asynchronous setting.
However, many of their results also hold or have obvious analogues in
McLean's synchronous setting (and ours hold in the asynchronous
setting).  The issue of synchrony
vs.~asynchrony is orthogonal to the issues we are discussing here.

Among other things, ZL also point out that
McLean's approach is insufficiently expressive.  In particular, they
focus on a property they call PSP (for {\em Perfect Security Property})
which they claim is not expressible using SIFs.%
\footnote{A proof of the result is sketched by Zakinthinos
\citeyear{Zakinthinos}. 
While we believe the claim, we suspect that a careful formal proof will be
much longer and more involved, in light of the difficulty of our own
proofs of Theorems~\ref{thm:notNOS} and~\ref{thm:DGNI}.}
They also introduce a general notion of security property that has some
of the flavor of McLean's notion of ``representable by SIFs'', in that
it is defined by a closure condition.
As in our approach, a security property 
for ZL
is a predicate on sets of systems.
However, for ZL, it is not an arbitrary predicate; it must satisfy an
additional constraint.
\dfn A predicate $\cS$ on $2^{\Sigma^*}$ is a {\em ZL-security property\/}
(on $\Sigma^*$) if there exists a predicate $Q$ on
$2^{\Sigma^*}$ such that, for all $\Sigma \subseteq \Sigma^*$,
$\cS({\Sigma})$ holds iff for all $\sigma \in {\Sigma}: \:
Q(LLES(\sigma,{\Sigma}))$ holds, where
$LLES(\sigma,\Sigma)=\{\tau |
\tau \in \Sigma \: \wedge \: {\tau}|_L={\sigma}|_L \} $ is the set of traces
with
the same low view as $\sigma$.
\edfn
That is, if a set $\Sigma$ of traces is in $\cS$, then for each trace in
$\Sigma$, $Q$ must hold for the
set of all traces in $\Sigma$ with the same low view as $\sigma$.
Conversely, if for each $\sigma \in \Sigma$, $Q$ holds for the set of
all traces in $\Sigma$ with the same low view as $\sigma$, then $\Sigma$
satisfies the security property.

It is not clear why this is a reasonable definition of ``security
property''.  There is certainly no independent motivation for it.
The following proposition gives at least one argument against it.

\pro The set of ZL-security properties is not closed under disjunction.
\epro
\prf Let $\Sigma^*$ consist of two traces, $\sigma_0$ and $\sigma_1$,
where the L's input and output are always 0 in $\sigma_0$ and always 1
in $\sigma_1$.  Thus, $LLES(\sigma_i, \Sigma^*) = \{\sigma_i\}$, for $i
= 0,1$.  Clearly $\cS_0 = \{\sigma_0\}$ and $\cS_1 = \{\sigma_1\}$ are
both ZL-security properties
(for $\cS_0$ we take $Q$ to hold on $\{\sigma_0\}$, while for $\cS_1$
we take $Q$ to hold on $\{\sigma_1\}$.)
However, $\cS_1 \union \cS_2$ is
not a ZL-security property.  For suppose it is; let $Q$ be the
corresponding security predicate.  Then both $Q(\{\sigma_0\})$ and
$Q(\{\sigma_1\})$ must hold.  But then $\Sigma^*$ would also satisfy
$\cS_1 \union \cS_2$, which it does not.  \eprf

On the other hand, Zakinthinos and Lee do show that a number of natural
security properties are ZL-security properties, including SEP and GNI.
A simple analysis shows that NOS is also a ZL-security property.
\pro NOS is a ZL-security property. \epro
\prf
It is easy to see that the definition of NOS is equivalent to the
following definition:
$$ NOS(\Sigma) \: \equiv \: \forall \sigma \in \Sigma \; \forall H \in
{\cal H} \; \exists \tau \in {\Sigma}_H \bigcap LLES(\sigma,\Sigma).$$
Let $ Q(A) \: \equiv \: \forall H \in {\cal H}. \; A \bigcap
{\Sigma}_H \neq \emptyset $.
Thus, $NOS(\Sigma) \:\equiv \: \forall \sigma \in \Sigma. \;
Q(LLES(\sigma,\Sigma)).$  \eprf

We now show that ZL-security properties are closed
under conjunction.  Since GNI is a ZL-security property, it follows that
DGNI is too.
\thm  The set of ZL-security properties is closed under conjunction.  \ethm

\prf Suppose $\cS$ and $\cS'$ are two
security properties with $Q$ and $Q'$ their corresponding security
predicates.
Then $\cS \wedge \cS'$ be the property $$\forall {\Sigma} \; \forall
\sigma
\in {\Sigma} \; (Q \wedge Q')(LLES(\sigma,\Sigma)).$$
It follows that $S \wedge S'$ is a security
property with corresponding security predicate $Q \wedge Q'$.
\eprf

As we said, ZL focus on a security property they call PSP.  To explain
PSP, we must first review the asynchronous systems considered by ZL.
For them (and also, for example, for Mantel \citeyear{Mantel00}), a
system is 
a tuple  $(E,I,O,\Sigma)$, where $E$ is a set of {\em events},
partitioned into two sets: $L$ and $H$ (low events and high events),
and $\Sigma$ is a set of traces, each of which is a finite
sequence of events in $E$.%
\footnote{Note that since ZL work in an asynchronous setting, their
notion of ``trace'' is different from that defined in
Section~\ref{sec:review}.  We continue to use the term ``trace''
even in the asynchronous setting, and hope that what we
mean is clear from context.}
Given a trace $\sigma$, let $\sigma_H$ denote the subsequence of
$\sigma$ consisting of the high events and let $\sigma_L$ denote the
subsequence consisting of low events.  It is quite straightforward to
reformulate notions like SEP, GNI, DGNI, and NOS in this framework; we
omit the details here.

The definition of PSP given by ZL is somewhat complicated.  Mantel
\citeyear{Mantel00} reformulates it in a  more comprehensible way.

\dfn A system $\Sigma$ satisfies PSP if and only if for all $\sigma \in
\Sigma$, 
$\sigma_L \in \Sigma$ and for 
all sequences of events $\alpha,\beta \in E^{*}$ and all events
$e \in E$, if $e \in H$, $\beta \alpha \in \Sigma$,  ${(\beta
\alpha)}_L={\sigma}_L$,
${\alpha}_H=\<\, \>$, and $\beta e \in \Sigma$, then it must be the case that
$\beta e \alpha \in \Sigma$.
\edfn

ZL show that PSP is a ZL-security property.  We show that it is also
representable by SIFs.

\pro PSP is representable by SIFs.
\epro

\prf
Let $F$ consist of the single SIF $f$, where $f(\sigma_1,\sigma_2) =
\beta e \alpha$ if there exist a high event $e \in H$ and sequences of
events $\alpha$ and $\beta$ such that ${\alpha}_H=\<\, \>$,
${\sigma}_1=\beta \alpha$,  and $\sigma_2 = \beta e$; otherwise
$f(\sigma_1,\sigma_2)=({\sigma_1})_L$.
Notice that $f$ is well defined since $\alpha$, $\beta$, and $e$, if
they exist, are uniquely determined by $\sigma_1$ and $\sigma_2$.
Notice also that $f(\sigma_1,\sigma_1)=({\sigma_1})_L$.

Suppose that $\Sigma$ satisfies PSP. Let ${\sigma}_1$ and
${\sigma}_2$ be two arbitrarily chosen traces in $\Sigma$. If
there exist a high event $e$ and  sequences of events $\alpha$ and
$\beta$ such 
that ${\sigma}_1=\beta \alpha$, ${\alpha}_H=\<\, \>$, and ${\sigma}_2=\beta
e$, then 
${(\beta \alpha)}_L=({\sigma_1})_L$, and PSP ensures that
$\beta e \alpha \in \Sigma$.
Since $f(\sigma_1, \sigma_2) = \beta e \alpha$ in this case,
$f(\sigma_1,\sigma_2) \in \Sigma$.  On the other hand, if there do not
exist such an $e$, $\alpha$, and $\beta$, then $f(\sigma_1,\sigma_2) =
({\sigma_1})_L \in \Sigma$ since $\Sigma$ satisfies PSP.
So $\Sigma$ is closed under $F$.

For the opposite implication, suppose  that $\Sigma$ is closed under
$F$. Suppose that $\sigma \in
\Sigma$, $\alpha, \beta \in E^*$, ${\alpha}_H=\<\, \>$, $e
\in H$,  $\beta \alpha \in \Sigma$, ${(\beta
\alpha)}_L={\sigma}_L$, and $\beta e \in \Sigma$.
Since $\Sigma$ is closed under $F$,
$f(\sigma_1, \sigma_2 ) = \beta e \alpha \in \Sigma$.
Also $f(\sigma_1,\sigma_1)=({\sigma_1})_L$, and so $({\sigma_1})_L \in \Sigma$.
But this
is exactly what we needed to prove that $\Sigma$
satisfies PSP.
\eprf

\section{Discussion}\label{sec:conc}
McLean's framework has
been the impetus for a number of frameworks for expressing security
properties (e.g., \cite{Mantel00,ZL97}), all based on defining
security properties in terms of closure conditions.
The question still remains as to what makes a framework ``good'' or
better than another.  Certainly one criterion is that an approach be
``natural'' and make it easy to express security properties.  Yet
another is that it be expressive, so that it can capture all natural
security properties.

We have examined McLean's SIF framework with regard to expressiveness.
Our results show that, as McLean presented it (considering only types),
the framework is insufficiently expressive to serve as a basis for
expressing security properties.  The fact that the properties
expressible are not closed under conjunction or disjunction, and natural
properties such as NOS and DGNI are not expressible, should suffice to
make that clear.  On the other hand, as we have shown, natural
extensions of the SIF framework are quite expressive.  In the process we
have shown that Zakinthinos and Lee's approach also has some
problems of expressibility; the set of security properties expressible
in their framework is not closed under disjunction.

\commentout{
We do have a translation of the closure under types condition into
Security Process Algebra (SPA), an extension of Milner's CCS proposed by
Focardi and al. \citeyear{FocardiSPA} to specify protocols with 2 levels
of clearance. We managed to prove that, with a proper definition of a
particular set of SPA processes called SPA McLean (SPAML), a system is
closed under a type if and only if for each pair of traces some relation
in the algebra holds (we do not present the  proofs here to keep the
dimensions of the paper within imposed limits).} 

The question still remains, of course, whether defining security
properties in terms of closure conditions is the way to go.
Mantel \citeyear{Mantel00} has perhaps the best-developed approach along these
lines.  He tries to provide a framework which
``provides the expressiveness of
Zakinthinos and Lee's framework with the elegance of McLean's''.
Certainly his  ``toolkit'' approach to
defining security properties seems promising.
Nevertheless, it is far from clear to us that basing a framework on
closure conditions is ultimately the right approach.
It would be interesting to compare the
expressive power and ease of use of these approaches to other
approaches, such as 
process algebra (see, for example, \cite{focardi01,ryan99,ryan00})
or a knowledge-based approach (see, for example,
\cite{BieberC92,HalOn02}).

\appendix
\section{Appendix: Proofs}\label{sec:proofs}
In this appendix, we prove Theorems~\ref{thm:notNOS}, \ref{thm:DGNI},
and \ref{disjTypes}.  We restate the theorems for the readers'
convenience.

\othm{thm:notNOS}
NOS is not $\cS_{\Hcal^*}$-representable by a type.
\eothm

\prf
We want to prove that there is no type $T$ such that 
for all systems $\Sigma$,
$\Sigma \in \Hcal^*$ satisfies NOS iff $\Sigma$ is closed under $T$.
As we observed, there are only 81 possible  types.
We proceed by a sequence of lemmas to eliminate each of these possibilities.
The first of these was already proved by McLean.
\lem\label{lem:SIM12} {\rm \cite[Theorem 2.4]{McLean94}} 
Let $T'$ be the result of replacing 1 by 2 and 2 by 1 in $T$.
(So, for example, if  $T$ is $T_{\langle (1:0),(2:1) \rangle}$,  
then $T'$ is $T_{\langle (2:0),(1:2)\rangle}$). Then
a system $\Sigma$ is closed under $T$ iff $\Sigma$ is closed under $T'$.
\elem

It is immediate from Lemma~\ref{lem:SIM12} that if there is a type $T$ that
represents NOS, then we can assume without loss of generality that 
$\mbox{{\em in}}^H \ne 2$.

The following lemma is straightforward, and is left to the reader.
\lem\label{lem:allsys}
All systems are closed under the following types: $T_{\langle
(0:0),(0:0) \rangle}$, $T_{\langle (0:0),(0:1) \rangle}$, $T_{\langle
(0:0),(1:0) \rangle}$, $T_{\langle (0:1),(0:0) \rangle}$, $T_{\langle
(1:0),(0:0) \rangle}$, $T_{\langle (0:0),(1:1) \rangle}$, $T_{\langle
(0:1),(0:1) \rangle}$, $T_{\langle (1:0),(0:1) \rangle}$, $T_{\langle
(1:0),(1:0) \rangle}$, $T_{\langle (0:1),(1:0) \rangle}$, $T_{\langle
(1:1),(0:0) \rangle}$, $T_{\langle (1:1),(1:0) \rangle}$, $T_{\langle
(1:1),(0:1) \rangle}$, $T_{\langle (1:0),(1:1) \rangle}$, $T_{\langle
(0:1),(1:1) \rangle}$, 
and $T_{\langle (1:1),(1:1) \rangle}$
(and their equivalent forms, as given by Lemma \ref{lem:SIM12}). 
\elem

Of course, it is immediate that if $\cS$ is a nontrivial security
property (i.e., $\cS \neq {\cS}^*$) then none of the types listed in
Lemma~\ref{lem:allsys} can represent $\cS$ (so, in particular, none of
them can represent NOS).

Consider now the system ${\Sigma}_{NOS}$ of Example~\ref{xam:NOSnotSEP} and recall that ${\Sigma}_{NOS}$ satisfies NOS.

\lem\label{lem:inH}
If $T=T_{\langle (in^H:in^L), (out^H:out^L)\rangle}$ ${\cS}_{{\cal H}^*}$-represents NOS, then $in^H=0$.
\elem
\prf
Suppose, by way of contradiction, that  $T=T_{\langle (in^H:in^L), (out^H:out^L)\rangle}$  ($in^H \neq 0$) ${\cS}_{{\cal H}^*}$-represents NOS. Based on Lemma~\ref{lem:SIM12}, with no loss of generality we can consider $in^H=1$. Since ${\Sigma}_{NOS}$ satisfies NOS, it means that  
it is closed under $T=T_{\langle (1:in^L), (out^H:out^L)\rangle}$. If at least one of $in^L, out^H, out^L$ is $2$, then the interleaving of $\sigma$ and $\tau$ results in a trace that, after the second step, has both zeros and ones, and so it is not in  ${\Sigma}_{NOS}$. So $in^L, out^H, out^L$ are either 0 or 1; but then, by Lemma~\ref{lem:allsys}, all systems are closed under $T$. This can't be true since NOS is not trivial.
\eprf

\lem\label{lem:inL}
If $T=T_{\langle (0:in^L), (out^H:out^L)\rangle}$ ${\cS}_{{\cal H}^*}$-represents NOS, then $in^L=0$.
\elem
\prf
Suppose, by way of contradiction, that $T=T_{\langle (0:in^L), (out^H:out^L)\rangle}$ with $in^L \neq 0$ ${\cS}_{{\cal H}^*}$-represents NOS. Based on Lemma~\ref{lem:SIM12}, with no loss of generality we can consider $in^L=1$. ${\Sigma}_{NOS}$ satisfies NOS, so it is closed under $T=T_{\langle (0:1), (out^H:out^L)\rangle}$. If at least one of $out^H$ and $out^L$ is $2$, then the interleaving of $\sigma$ and $\tau$ contains both  0 and 1 after the second step, and so it is not in ${\Sigma}_{NOS}$. Then $out^H$ and $out^L$ are both 0 or 1; by Lemma~\ref{lem:allsys}, all systems are closed under $T$, which contradicts the fact that NOS is not trivial.
\eprf

Following the same pattern, we can prove 
\lem\label{lem:outH}
If $T=T_{\langle (0:0), (out^H:out^L)\rangle}$ ${\cS}_{{\cal H}^*}$-represents NOS, then $out^H=0$.
\elem
 
From Lemmas~\ref{lem:inH},~\ref{lem:inL} and \ref{lem:outH} it follows
that $T=T_{\langle (0:0), (0:out^L)\rangle}$, but then by
Lemma~\ref{lem:allsys} and since NOS is not trivial, $T$ cannot
${\cS}_{{\cal H}^*}$-represent NOS.  
\eprf

\medskip
\othm{thm:DGNI}
DGNI is not ${\cS}^*$-representable by 
a type.
\eothm

\prf
Suppose, by way of contradiction, that there is a type $T=T_{\langle (in^H:in^L),(out^H:out^L)\rangle}$ that ${\cS}^*$-represents DGNI. The following two lemmas establish a contradiction:
\lem
If $T=T_{\langle (in^H:in^L),(out^H:out^L)\rangle}$ ${\cS}^*$-represents DGNI, then at least one of $in^H, in^L, out^H, out^L$ is 0.
\elem
\prf
Recall ${\Sigma}_{DGNI}$ of Example~\ref{xam:DGNInotSEP} with 15 traces
of the form ${(x_1,x_2,x_3,x_4)}^\omega$, $x_1$, $x_2$, $x_3$ and $x_4$ 0 or
1, with the exception of ${(1,0,1,0)}^\omega$. ${\Sigma}_{DGNI}$ satisfies
DGNI. If $T=T_{\langle (in^H:in^L),(out^H:out^L)\rangle}$
${\cS}^*$-represents DGNI, then ${\Sigma}_{DGNI}$ is closed under $T$.  

Suppose, by way of contradiction, that none of $in^H$, $in^L$, $out^H$ and $out^L$ is 0. By Lemma~\ref{lem:SIM12}, we can assume with no loss of generality that $in^H=1$. If all $in^L$, $out^H$ and $out^L$ are 1, then by Lemma~\ref{lem:allsys}, all systems are closed under $T$, which contradicts the fact that NOS is not trivial. So at least one of $in^L$, $out^H$ and $out^L$ is 2. Take $\tau={(0,0,1,0)}^\omega$; $\tau \in {\Sigma}_{DGNI}$. Take $\sigma={(1,x,y,z)}^\omega$ obtained from ${(1,0,1,0)}^\omega$ in the following way: if $in^L=1$ take $x=0$, otherwise take $x=1$; if $out^H=1$ then take $y=1$, otherwise $y=0$; if $out^L=1$ take $z=0$, otherwise $z=1$. Since at least one of $in^L$, $out^H$ and $out^L$ is 2, $\sigma \in {\Sigma}_{DGNI}$. But an interleaving of type $T$ of $\sigma$ and $\tau$ results into ${(1,0,1,0)}^\omega$, which is not in ${\Sigma}_{DGNI}$. This contradicts the fact that ${\Sigma}_{DGNI}$ is closed under $T$.
\eprf

\lem
If $T=T_{\langle (in^H:in^L),(out^H:out^L)\rangle}$ ${\cS}^*$-represents
DGNI, then none of $in^H, in^L, out^H, out^L$ is 0. 
\elem
\prf
Consider the system $\Sigma_{notGNI}$ with 8 traces of the form ${(x_1,x_1,x_2,x_3)}^\omega$, $x_1,x_2,x_3 \in \{ 0,1\}$. $\Sigma_{notGNI}$ does not satisfy GNI, since an interleaving of type $T_{\langle (1:2),(0:2) \rangle}$ of traces ${\sigma}_1={(0,0,0,0)}^\omega$ and ${\sigma}_2={(1,1,1,1)}^\omega$, both in  $\Sigma_{notGNI}$, has the form ${(0,1,x,1)}^\omega$, which is not in $\Sigma_{notGNI}$. It follows that $\Sigma_{notGNI}$ does not satisfy DGNI, too. $\Sigma_{notGNI}$ is closed under all types $T=T_{\langle (in^H:in^L),(out^H:out^L)\rangle}$ with $in^H=0$ or $in^L=0$; it follows that if $T=T_{\langle (in^H:in^L),(out^H:out^L)\rangle}$ ${\cS}^*$-represents DGNI, then $in^H \neq 0$ and $in^L \neq 0$.

Consider the system ${\Sigma}_{GNInotDGNI}$ consisting of 8 traces of the form ${(x_1,x_2,x_2,x_3)}^\omega$, with $x_1,x_2,x_3 \in \{ 0,1\}$. ${\Sigma}_{GNInotDGNI}$ satisfies GNI, since an interleaving of type $T_{\langle (1:2),(0:2) \rangle}$ of two traces ${(x_1,x_2,x_2,x_3)}^\omega$ and ${(y_1,y_2,y_2,y_3)}^\omega$ has the form ${(x_1,y_2,x,y_3)}^\omega$, and for $x=y_2$ this is a trace in ${\Sigma}_{GNInotDGNI}$. But ${\Sigma}_{GNInotDGNI}$ does not satisfy reverse GNI, and for this reason DGNI too, since an interleaving of type $T_{\langle (1:2),(1:0) \rangle}$ of traces $({0,0,0,0)}^\omega$ and ${(1,1,1,1)}^\omega$ has the form ${(0,1,0,x)}^\omega$, which is not in ${\Sigma}_{GNInotDGNI}$. ${\Sigma}_{GNInotDGNI}$ is closed under all types $T=T_{\langle (in^H:in^L),(out^H:out^L)\rangle}$ with $in^L=0$ or $out^H=0$. It means that, if $T=T_{\langle (in^H:in^L),(out^H:out^L)\rangle}$ ${\cS}^*$-represents DGNI, then $in^L \neq 0$ and $out^H \neq 0$.

Finally, take ${\Sigma'}_{notGNI}$ to be the system with 8 traces of the form ${(x_1,x_2,x_3,x_1)}^\omega$, $x_1, x_2, x_3 \in \{0,1\}$; ${\Sigma'}_{notGNI}$ does not satisfy GNI, and for this reason DGNI either, since an interleaving of type $T_{\langle (1:2),(0:2)\rangle}$ of ${(0,0,0,0)}^\omega$ and ${(1,1,1,1)}^\omega$, both in ${\Sigma'}_{notGNI}$, has the form ${(0,1,x,1)}^\omega$, which is not in ${\Sigma'}_{notGNI}$. ${\Sigma'}_{notGNI}$ is closed under all types $T=T_{\langle (in^H:in^L),(out^H:out^L)\rangle}$ with $in^H=0$ or $out^L=0$. It follows that, if $T=T_{\langle (in^H:in^L),(out^H:out^L)\rangle}$ ${\cS}^*$-represents DGNI, then $in^H \neq 0$ and $out^L \neq 0$.
\eprf
\medskip

\othm{disjTypes}
The set of security properties representable by types is not closed
under disjunction.
\eothm

\prf
The proof is a corollary of the following proposition:
\pro\label{pro:lackDisj}
Let $\cS$ be the security property represented by $T_{\langle (1:2),(2:2)\rangle}$, and $\cS'$ the security property resulting from the disjunction of SEP and $\cS$. Then  $\cS'$ is not ${\cS}^*$-representable by types.
\epro
\prf
Suppose, by way of contradiction, that there is some type
$T=T_{\langle (in^H:in^L),(out^H:out^L) \rangle}$ that represents
$\cS'$. 
Then a system is closed under $T_{\langle
(1:2),(1:2) \rangle}$ (the type corresponding to SEP) or $T_{\langle 
(1:2),(2:2) \rangle}$ if and only if it is  closed under  $T$. 
Let ${\Sigma}_{SEP}$ be the system consisting of the 8 traces of the
form ${(x_1,x_2,x_3,x_2)}^\omega$, with $x_1$, $x_2$, and $x_3$ $\in
\{0,1\}$.  Thus, in all traces of ${\Sigma}_{SEP}$, the low output is
the same as the low input and independent of the high view. So
${\Sigma}_{SEP}$ satisfies SEP. 
It is easy to see that both 
$\Sigma_{GNInotDGNI}$ and $\Sigma_{SEP}$ 
are in $\cS'$, since $\Sigma_{GNInotDGNI}$ is closed under $T_{\langle
(1:2),(2:2)\rangle}$ and 
$\Sigma_{SEP}$ satisfies SEP.
It is also easy to see that neither
$\Sigma_{notGNI}$ nor ${\Sigma'}_{notGNI}$ is in  $\cS'$.
$\Sigma_{SEP}$ satisfies SEP.
since neither system satisfies SEP and neither is closed under
$T_{\langle (1:2),(2:2)\rangle}$. 
From Lemma~\ref{lem:SIM12}, it follows that
there is a type $T_{\langle (in^H:in^L),(out^H:out^L)\rangle}$ that
${\cS}^*$-represents $\cS'$ if and only if  there is a type $T_{\langle
(in'^H:in'^L),(out'^H:out'^L)\rangle}$ with $in'^H \neq 2$ that
${\cS}^*$-represents $\cS'$. 
Thus, it suffices to show that there is no type that represents $\cS'$
that has $in^H$ being 0 or 1.
The following two lemmas show that neither case can happen.
\lem
There is no type $T=T_{\langle (0:in^L),(out^H:out^L)\rangle}$ that ${\cS}^*$-represents $\cS'$.
\elem
\prf
Suppose, by of contradiction, that $T=T_{\langle
(0:in^L),(out^H:out^L)\rangle}$ ${\cS}^*$-represents $\cS'$. 
All systems are closed under $T_{\langle (0:0), (out^H:out^L)\rangle}$
for $\langle out^H, out^L\rangle \not \in \{\langle 1,2\rangle, \langle
2,1\rangle\}$ and under $T_{\langle (0:2),(0:out^L\rangle}$ for $out^L
\in \{ 0,2\}$. Since $\cS'$ is not trivial, we  
can rule out all these types.
By Lemma\ref{lem:SIM12}, type $T=T_{\langle (0:0),(1:2)\rangle}$ is
equivalent to $T_{\langle (0:0),(2:1)\rangle}$, and  
$\Sigma_{notGNI}$ is closed under $T$, although it is not in $\cS'$;
similarly, 
$\Sigma_{GNInotDGNI}$ is not closed under $T$, but is not in $\cS'$.
\eprf
\lem
There is no type $T=T_{\langle (1:in^L),(out^H:out^L)\rangle}$ that ${\cS}^*$-represents $\cS'$.
\elem
\prf
Again, suppose by way of contradiction that there is some type
$T=T_{\langle (1:in^L),(out^H:out^L)\rangle}$ that ${\cS}^*$-represents
$\cS'$. 
If $in^L \in \{0,1\}$, then 
$\Sigma_{notGNI}$ is closed under $T$, but it is not in $\cS'$. 
${\Sigma'}_{notGNI}$ is closed under $T_{\langle
(1:2),(out^H:1)\rangle}$, $T_{\langle (1:2),(2:0)\rangle}$ and
$T_{\langle (1:2),(0:0)\rangle}$, but is not in $\cS'$, so we can rule
out these types too. $T$ cannot be any of the types $T_{\langle
(1:2),(1:out^L)\rangle}$ since  
$\Sigma_{GNInotDGNI} \in \cS'$ and is not closed under them; similarly,
$T \neq T_{\langle (1:2),(2:2)\rangle}$ since  
$\Sigma_{SEP}$ is not closed under it, while it is in $cS'$.

We are left with the type $T_{\langle (1:2),(0:2)\rangle}$ that
represents GNI. 
\commentout{It is not difficult to show that this
type is at most as general as $T_{\langle (1:2),(0:2) \rangle}$, which
is actually  the type representing  GNI. }
Consider the system $\Sigma$ with 8 traces of the form
${(0,x_1,x_1,x_2)}^\omega$ 
and 8 traces of the form  
${(1,x_1,1-x_1,x_2)}^\omega$, $x_1, x_2 \in \{0,1\}$. 
$\Sigma$
satisfies GNI, since any low view is compatible with any
high input sequence. 
Thus, $\Sigma$ is closed under the type $T_{\langle
(1:2),(0:2)\rangle}$. 
However, $\Sigma$ is not separable and it is not in $\cS$, hence $\Sigma
\notin \cS'$.  Thus, 
$T_{\langle (1:2),(0:2)\rangle}$ does not ${\cS}^*$-represent $\cS'$.
\eprf

\bibliography{z,joe}
\bibliographystyle{chicago}

\end{document}

%% file: defn.tex
\newtheorem{THEOREM}{Theorem}[section]
\newenvironment{theorem}{\begin{THEOREM} \hspace{-.85em} {\bf :} }%
                        {\end{THEOREM}}
\newtheorem{LEMMA}[THEOREM]{Lemma}
\newenvironment{lemma}{\begin{LEMMA} \hspace{-.85em} {\bf :} }%
                      {\end{LEMMA}}
\newtheorem{COROLLARY}[THEOREM]{Corollary}
\newenvironment{corollary}{\begin{COROLLARY} \hspace{-.85em} {\bf :} }%
                          {\end{COROLLARY}}
\newtheorem{PROPOSITION}[THEOREM]{Proposition}
\newenvironment{proposition}{\begin{PROPOSITION} \hspace{-.85em} {\bf :} }%
                            {\end{PROPOSITION}}
\newtheorem{DEFINITION}[THEOREM]{Definition}
\newenvironment{definition}{\begin{DEFINITION} \hspace{-.85em} {\bf :} \rm}%
                            {\end{DEFINITION}}
\newtheorem{CLAIM}[THEOREM]{Claim}
\newenvironment{claim}{\begin{CLAIM} \hspace{-.85em} {\bf :} \rm}%
                            {\end{CLAIM}}
\newtheorem{EXAMPLE}[THEOREM]{Example}
\newenvironment{example}{\begin{EXAMPLE} \hspace{-.85em} {\bf :} \rm}%
                            {\end{EXAMPLE}}
\newtheorem{REMARK}[THEOREM]{Remark}
\newenvironment{remark}{\begin{REMARK} \hspace{-.85em} {\bf :} \rm}%
                            {\end{REMARK}}

\newcommand{\thm}{\begin{theorem}}
\newcommand{\lem}{\begin{lemma}}
\newcommand{\pro}{\begin{proposition}}
\newcommand{\dfn}{\begin{definition}}
\newcommand{\rem}{\begin{remark}}
\newcommand{\xam}{\begin{example}}
\newcommand{\cor}{\begin{corollary}}
\newcommand{\prf}{\noindent{\bf Proof:} }
\newcommand{\ethm}{\end{theorem}}
\newcommand{\elem}{\end{lemma}}
\newcommand{\epro}{\end{proposition}}
\newcommand{\edfn}{\bbox\end{definition}}
\newcommand{\erem}{\bbox\end{remark}}
\newcommand{\exam}{\bbox\end{example}}
\newcommand{\ecor}{\end{corollary}}
\newcommand{\eprf}{\bbox\vspace{0.1in}}
\newcommand{\beqn}{\begin{equation}}
\newcommand{\eeqn}{\end{equation}}

\newcommand{\bbox}{\vrule height7pt width4pt depth1pt}

\newcommand{\clm}{\begin{claim}}
\newcommand{\eclm}{\end{claim}}

\newcommand{\union}{\cup}
\newcommand{\inter}{\cap}

\renewcommand{\phi}{\varphi}

\newcommand{\Hcal}{{\cal H}}
\renewcommand{\P}{{\cal P}}

\newcommand{\Z}{{\cal Z}}

\newcommand{\<}{\langle}
\renewcommand{\>}{\rangle}

\newcommand{\ol}{\setlength{\itemsep}{0pt}\begin{enumerate}}
\newcommand{\eol}{\end{enumerate}\setlength{\itemsep}{-\parsep}}
\newcommand{\ul}{\setlength{\itemsep}{0pt}\begin{itemize}}
\newcommand{\dl}{\setlength{\itemsep}{0pt}\begin{description}}
\newcommand{\edl}{\end{description}\setlength{\itemsep}{-\parsep}}
\newcommand{\eul}{\end{itemize}\setlength{\itemsep}{-\parsep}}

\newcommand{\cS}{{\cal S}}

\newcommand{\commentout}[1]{}

\newcommand{\bi}{\begin{itemize}}
\newcommand{\ei}{\end{itemize}}
\newcommand{\be}{\begin{enumerate}}
\newcommand{\ee}{\end{enumerate}}